\documentclass[sigconf]{acmart}
\usepackage{amsmath,nccmath,tabularx} 
\usepackage{algorithm}
\usepackage{algorithmic}
\usepackage{graphicx}
\graphicspath{{images/}}
\usepackage{wrapfig,lipsum}
\usepackage{tabularx}
\usepackage{subcaption}
\usepackage{multirow}

\usepackage{enumitem}
\setlist{leftmargin=0mm}
\usepackage[T1]{fontenc}
\usepackage[utf8]{inputenc}
\usepackage[font=small,labelfont=bf]{caption}

\usepackage{listings}
\usepackage{balance}

\usepackage{xcolor}

\colorlet{punct}{red!60!black}
\definecolor{background}{HTML}{EEEEEE}
\definecolor{delim}{RGB}{20,105,176}
\colorlet{numb}{magenta!60!black}
\lstdefinelanguage{json}{
	basicstyle=\normalfont\ttfamily,
	numberstyle=\scriptsize,
	stepnumber=1,
	showstringspaces=false,
	literate=
	*{0}{{{\color{numb}0}}}{1}
	{1}{{{\color{numb}1}}}{1}
	{2}{{{\color{numb}2}}}{1}
	{3}{{{\color{numb}3}}}{1}
	{4}{{{\color{numb}4}}}{1}
	{5}{{{\color{numb}5}}}{1}
	{6}{{{\color{numb}6}}}{1}
	{7}{{{\color{numb}7}}}{1}
	{8}{{{\color{numb}8}}}{1}
	{9}{{{\color{numb}9}}}{1}
	{:}{{{\color{punct}{:}}}}{1}
	{,}{{{\color{punct}{,}}}}{1}
	{\{}{{{\color{delim}{\{}}}}{1}
	{\}}{{{\color{delim}{\}}}}}{1}
	{[}{{{\color{delim}{[}}}}{1}
	{]}{{{\color{delim}{]}}}}{1},
}

\AtBeginDocument{%
  \providecommand\BibTeX{{%
    \normalfont B\kern-0.5em{\scshape i\kern-0.25em b}\kern-0.8em\TeX}}}

\copyrightyear{2022} 
\acmYear{2022} 
\setcopyright{acmlicensed}\acmConference[SIGIR '22]{Proceedings of the 45th International ACM SIGIR Conference on Research and Development in Information Retrieval}{July 11--15, 2022}{Madrid, Spain}
\acmBooktitle{Proceedings of the 45th International ACM SIGIR Conference on Research and Development in Information Retrieval (SIGIR '22), July 11--15, 2022, Madrid, Spain}
\acmPrice{15.00}
\acmDOI{10.1145/3477495.3531658}
\acmISBN{978-1-4503-8732-3/22/07}
\begin{document}
\fancyhead{}
\title{Asyncval: A Toolkit for Asynchronously Validating Dense Retriever Checkpoints during Training}


\author{Shengyao Zhuang}
\affiliation{%
	\institution{The University of Queensland}
	\streetaddress{4072 St Lucia}
	\city{Brisbane}
	\state{QLD}
	\country{Australia}}
\email{s.zhuang@uq.edu.au}

\author{Guido Zuccon}
\affiliation{%
	\institution{The University of Queensland}
	\streetaddress{4072 St Lucia}
	\city{Brisbane}
	\state{QLD}
	\country{Australia}}
\email{g.zuccon@uq.edu.au}


\begin{abstract}
The process of model checkpoint validation refers to the evaluation of the performance of a model checkpoint executed on a held-out portion of the training data while learning the hyperparameters of the model. This model checkpoint validation process is used to avoid over-fitting and determine when the model has converged so as to stop training.
A simple and efficient strategy to validate deep learning checkpoints is the addition of validation loops to execute during training. However, the validation of dense retrievers (DR)  checkpoints is not as trivial -- and the addition of validation loops is not efficient.
This is because, in order to accurately evaluate the performance of a DR checkpoint, the whole document corpus needs to be encoded into vectors using the current checkpoint before any actual retrieval operation for checkpoint validation can be performed. This corpus encoding process can be very time-consuming if the document corpus contains millions of documents (e.g., 8.8M for MS MARCO v1 and 21M for Natural Questions). Thus, a na\"ive use of validation loops during training will significantly increase training time. To address this issue, we propose \textit{Asyncval}: a Python-based toolkit for efficiently validating DR checkpoints during training. Instead of pausing the training loop for validating DR checkpoints, Asyncval decouples the validation loop from the training loop, uses another GPU to automatically validate new DR checkpoints and thus permits to perform validation  asynchronously from training. 
Asyncval also implements a range of different corpus subset sampling strategies for validating DR checkpoints; these strategies allow to further speed up the validation process. We provide an investigation of these methods in terms of their impact on validation time and validation fidelity.
Asyncval is made available as an open-source project at \url{https://github.com/ielab/asyncval}.

\end{abstract}

\ccsdesc[500]{Information systems~Evaluation of retrieval results}
\keywords{dense retriever, checkpoints validation, neural information retrieval}

\maketitle

\section{Introduction} \label{sec:intro}
Pre-trained language models (PLMs) based neural ranking models have shown promising effectiveness on information retrieval tasks~\cite{lin2021pretrained} -- in this paper we specifically focus on passage retrieval, although methods, tool and findings can be extended to documents. Among these models, a great effort has been devoted to developing effective dense retrievers (DRs)~\cite{karpukhin-etal-2020-dense,xiong2020approximate,gao2021complementing,zhan2020repbert,khattab2020colbert,DBLP:conf/emnlp/GaoC21,gao2021unsupervised,Zhan2021OptimizingDR,ren2021pair,ren2021rocketqav2,qu2021rocketqa,hofstatter2020improving,zhuang2021dealing,zhuang2022characterbert,yu2021improving,li2022improving,li2021pseudo}. DRs exploit a PLM-based bi-encoder architecture to separately encode queries and passages into low-dimensional dense vectors. Then the passage vectors will be indexed and the retrieval is performed by estimating the similarity of the query vector against all the passage vectors in the index. 

At first sight, the task a DR is trained for is similar to the sentence similarity task in natural language processing in that the inputs of the model are two pieces of text and the goal of the model is to estimate the similarity of the two texts~\cite{lin2021proposed}. One key difference, however, is that DRs are to be exploited in a retrieval pipeline and thus they are required to encode all the passages in the corpus into dense vectors. 
It is important to highlight this difference because, as we shall see below, it is this difference that poses a key efficiency challenge when validating DR model checkpoints during training -- and this challenge is not present in tasks such that of training (with validation) a model for the sentence similarity task. With validation of a DR model checkpoint during training we refer to the process of evaluating the effectiveness of a checkpoint training as a way to guide decisions about model convergence and training termination.
In this demo paper we present a method and its corresponding implementation into a usable tool for validating DRs in an efficient manner.

So, how are deep learning models commonly validated during training?
The most common (and straightforward) validation approach is the use, during training, of  a held-out validation set to evaluate some of the checkpoints' performance on the task (e.g., classification accuracy for the sentence classification task). 
Checkpoints to validate are typically selected according to a fix step schedule (e.g., a validation every $k$ thousand steps) or at the end of each training epoch. 
Once a checkpoint has been constructed and validation time is reached, the training process is stopped, the model parameters are fixed and saved, and inference on the validation set is performed and the corresponding effectiveness on the target task (e.g., retrieval effectiveness for DRs) or loss on validation set is recorded.
Researchers then can select the best model checkpoint among the set of available, validated checkpoints, or track the validation effectiveness to decide e.g., when it plateaus and training can be stopped. This validation approach forms the cornerstone of early-stopping techniques, which have been shown effective in preventing model over-fitting~\cite{prechelt1998early}. 


\begin{figure*}
		\begin{subfigure}{1\linewidth}
		\centering
		\includegraphics[width=\linewidth]{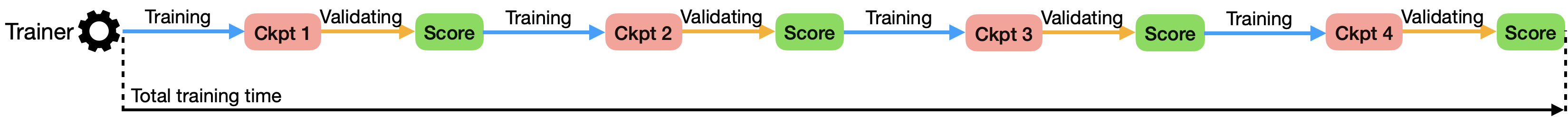}
		\caption{Standard validation.}
		\label{fig:sfig1}
	\end{subfigure}%
\\
	\begin{subfigure}{1\linewidth}
		\centering
		\includegraphics[width=\linewidth]{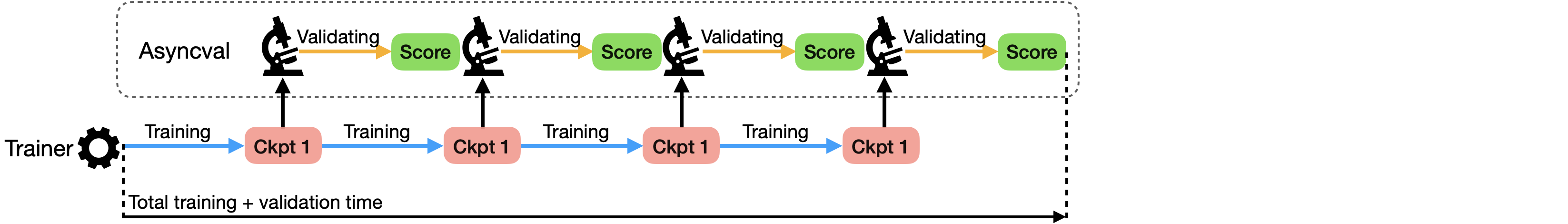}
		\caption{Asyncval validation.}
		\label{fig:sfig2}
	\end{subfigure}
	\caption{Standard validation VS Asyncval validation for validating dense retriever model checkpoints during training. }
	\label{fig:fig1}
\end{figure*}

However, this simple validation approach is not suitable for validating DR model checkpoints. This is because although retrieval based on dense vectors  is fast and quickly scales to large amounts of queries, the DR-based system requires all  passages in the corpus be first encoded into dense vectors, before the actual retrieval can take place. 
Although the time complexity of the corpus encoding process being linear with the corpus size, the process itself is very time consuming because passage corpora are often very large and  pre-trained language models are also large. 
Let us consider for example the training if a BERT-based DR with the MS MARCO passage ranking dataset~\cite{nguyen2016ms}, which contains $\approx 8.8$M passages: the corpus encoding process on a powerful GPU (E.g. Nvidia V100) can take up to 2 hours. Since the standard validation approach requires this encoding process to be repeated for $n$ epochs or training steps, the total time spent on re-encoding repetitively MS MARCO with a new DR checkpoint is $n \times 2 \mbox{hours}$. Figure~\ref{fig:sfig1}(top) illustrates an example of validating $n= 4$ checkpoints during DR training using the standard validation approach: the validation alone would thus take in excess of 8 hours ($n \times 2$ hours for corpus encoding plus any time spent in actually performing the retrieval process on the validation query set).


In an effort to speed up DR checkpoints validation during training, and thus reduce the overall time taken by the training with validation process, we developed \texttt{Asyncval}, a Python-based dense retrievers validation toolkit. 
Asyncval uncouples the validation process from the main training routine; this is done such that corpus encoding and validation query set retrieval and evaluation can be executed asynchronously from the training process.
By doing so, training and validation can be performed in parallel (Figure~\ref{fig:sfig2}). 
This has several implications. First,  it means that in Asyncval, the validation of the previous checkpoint is running while the next checkpoint is being trained. Second, the total training and validation time in Asyncval is equal to the sum of training time of each checkpoint plus the last checkpoint validation time, if each validation time is less than each training time; alternatively, the total training time is equal to the first checkpoint's training time plus the sum of each validation time (if validation time is more than training time) -- Figure~\ref{fig:fig1} bottom. This is unlike in the common training and validation approach where the total training time is equal to the sum of the training time and validation time of each checkpoint (Figure~\ref{fig:fig1} top).
Hence Asyncval can potentially save a large amount of the total time required to train and validate a DR. 
In addition, Asyncval provides a corpus subset sampling feature that allows to select and only encode a subset of the corpus based on the results of a given baseline. We show that a checkpoint's validation time can be reduced from 2 hours to 10 minutes with high \textit{validation fidelity}, i.e. the validation results on these subsets highly align with the results obtained on the whole validation corpus.

\section{Related Works} \label{sec:related_work}

Model validation during training is an aspect rarely detailed in the dense retrievers literature -- often papers just report the epoch or training step used to select the model checkpoint for further evaluation on the test data. Some exceptions do exist. The authors of ANCE report selecting the checkpoint based on the training loss curve or otherwise by the loss on the validation set~\cite{xiong2020approximate}.
They further report that the training loss is well aligned with the retrieval effectiveness -- based on this they decided not to rely on the entire validation query set to select the model checkpoint.

A different validation strategy was proposed alongside DPR~\cite{karpukhin-etal-2020-dense}~\footnote{See https://github.com/facebookresearch/DPR}.
Instead of using training loss, they use average rank validation, where the passages from the input data and the gold passages of each validation query are aggregated into a pool.
Then the average rank of the gold passage for each validation query is used as an indicator of the checkpoints' final retrieval performance. By doing this, there is no need to encode the whole corpus:  only a small subset of the corpus needs to be encoding, thus speeding up validation. They found that this validation strategy is more correlated with the final retrieval performance than loss-based validation. Our Asyncval also adopts this strategy as one of the features it offers to speed up DR checkpoints validation. 

Rapid validation of DR checkpoints can be obtained also by considering the passage re-ranking task, rather than the full retrieval task. This means that instead of encoding the whole corpus, one can just encode the top $k$ passages of each validation query. This approach is used by RocketQA~\cite{qu2021rocketqa}\footnote{Information obtained from a discussion with the original authors.}.

\section{Asyncval design philosophy} \label{sec:approach}

The design of Asyncval aimed to achieve the following goals:
\begin{itemize}[leftmargin=*]
	\item Ease of use: we wanted that the use of Asyncval can be easily integrated into the training loop of new dense retrievers -- regardless of the DR architecture or dataset used.
	
	\item Fast validation: we wanted that validation is performed in Asyncval as quickly as possible, so as to embed validation in the training practice, without major overheads.
	
	\item High fidelity: we wanted that the validation insights obtained with Asyncval are a close resemblance of the final full retrieval score. 
\end{itemize}

For this, we designed Asyncval to be a closed-loop, i.e. researchers do not need to implement the main validation process; to use Asyncval they instead just need to provide a corpus file and a validation query set file and override an encoder abstract Python class.

Corpus and validation query set files are to be in JSON format; specifically each line in the file is a JSON object that represents a passage or a query:
\begin{lstlisting}[language=json,firstnumber=1]
{"text_id": str, "text": List[int]}
\end{lstlisting}
where ``text\_id'' is a unique id for a passage or query in the corpus or query file, and ``text'' is a list of integers representing the token ids, including ids for the tokens in the passage or query text and any special token ids (such as [CLS] and [SEP]). There are two reasons for requiring Asyncval's users to tokenize their passages and queries and thus supply only token ids:  (1) Different DRs may use different customized tokenizers and special tokens, (2) Pre-tokenizing all text at once can speed up validation as there is no need to tokenize the same query and passage for each model checkpoint.


After the corpus and query file are prepared, a python class called \texttt{DenseModel} needs to be defined; this class inherits from our \texttt{asyncval/modelling.py}:

\begin{figure*}[t!]
	\includegraphics[width=0.95\columnwidth]{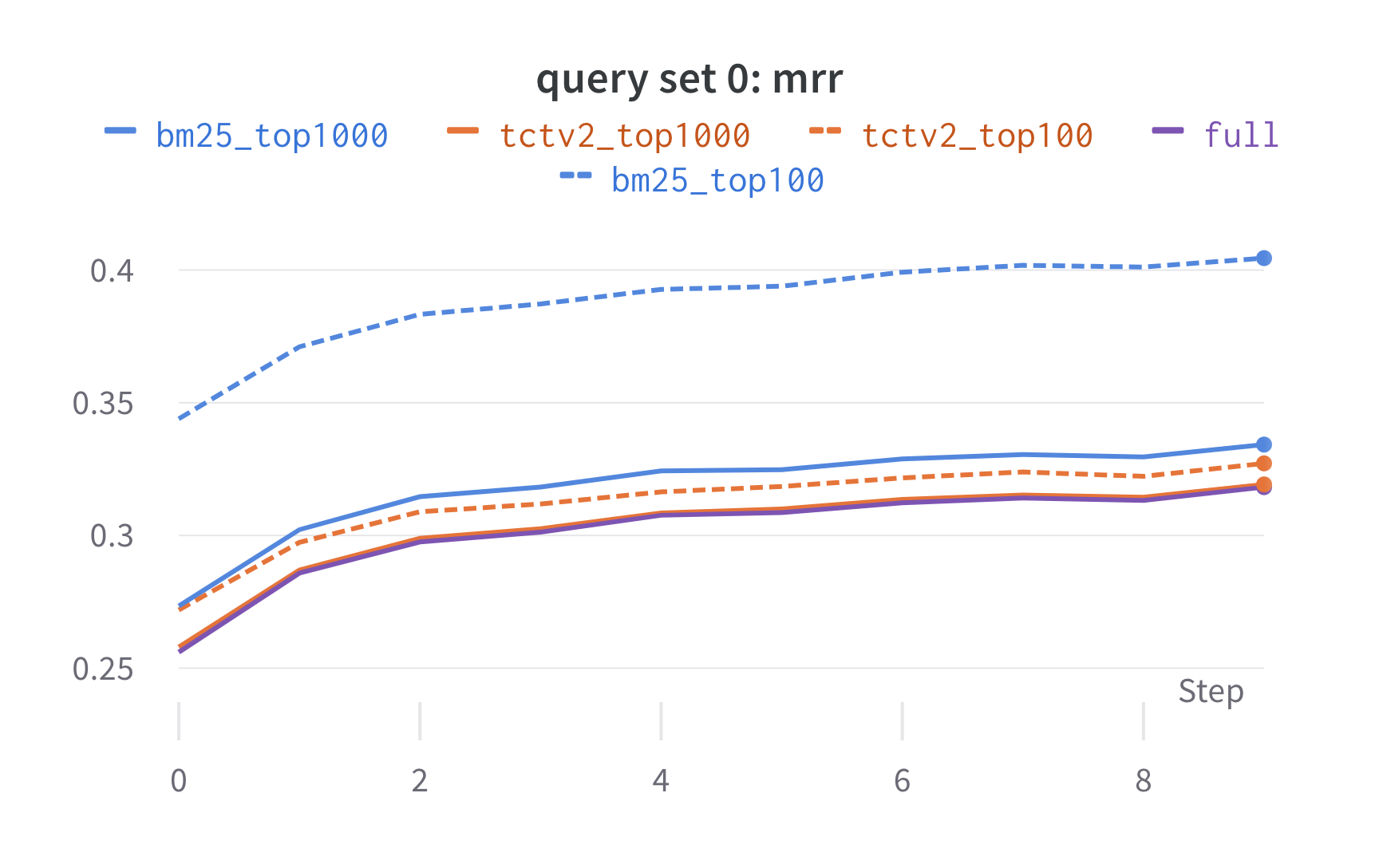}
	\includegraphics[width=0.95\columnwidth]{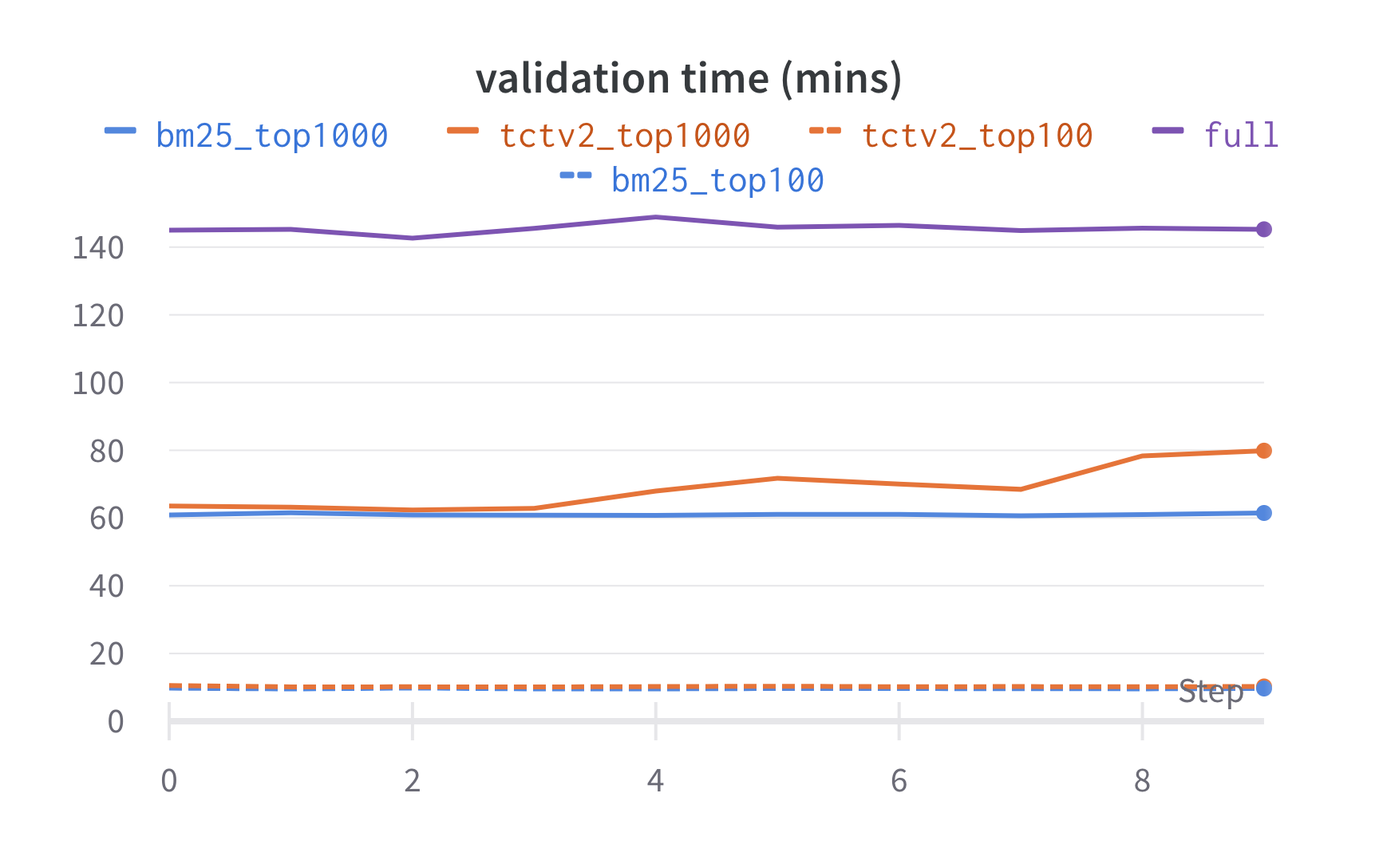}
	\caption{Wandb dashboard with Asyncval. Left: MRR@10 for each checkpoint. Right: the time required to validate each checkpoint. }
	\label{fig:fig2}
\end{figure*}

\begin{lstlisting}[language=python, label={code:avg}]
class Encoder(torch.nn.Module):
	def __init__(self, ckpt_path: str, async_args: AsyncvalArguments):
		super().__init__()
		self.ckpt_path = ckpt_path
		self.async_args = async_args
		
	@torch.no_grad()
	def encode_passage(self, psg: Dict[str, Tensor]) -> Tensor:
		raise NotImplementedError

	@torch.no_grad()
	def encode_query(self, qry: Dict[str, Tensor]) -> Tensor:
		raise NotImplementedError
\end{lstlisting}
The \texttt{Encoder} class is a \texttt{torch.nn.Module} and it only has three methods. The constructor method only takes the checkpoint path and the pre-defined \texttt{AsyncvalArguments} as inputs and creates the corresponding in-class variables. In \texttt{AsyncvalArguments} we pre-defined several useful arguments such as the maximum query and document lengths that can be passed via command line. Researchers can use these variables to easily build their DR model in the \texttt{DenseModel} subclass. The other two methods, namely \texttt{encode\_query} and \texttt{encode\_passage}, are responsible to encode queries and passages into dense vectors (represented by torch \texttt{Tensor}). These two methods take a dictionary of key and values as input, where keys are Huggingface transformers~\cite{wolf-etal-2020-transformers} input keys such as \texttt{input\_ids} and \texttt{attention\_mask} and values are batched token ids and an attention mask matrix. Researchers need to implement and override these two methods by (i) taking the input dictionary and feed it to the DR model that they built in the constructor method, and (ii) returning batch dense vectors. 

Finally, after the \texttt{DenseModel} and the corpus and query files ave been prepared, researchers can simply run Asyncval with the following command:
\begin{lstlisting}[language=bash, label={code:avg}]
python -m asyncval \
		--query_file List[str] \
		--candidate_dir str \
		--ckpts_dir str \
		--tokenizer_name_or_path str \
		--q_max_len int \
		--p_max_len int \
		--qrel_file str \
		--run_name str \
		--write_run bool \
		--output_dir str \
		--max_num_valid int \
		--logging_dir str \
		--metrics List[str] \
		--report_to List[str]
\end{lstlisting}
Arguments usage is as follows:
\begin{itemize}[leftmargin=*]
	\item \texttt{-query\_file}: the path to the pre-tokenized query JSON file. Multiple files can be provided;
	\item \texttt{-candidate\_dir}: the path to the folder where the pre-tokenized corpus JSON files are saved;
	\item \texttt{-ckpts\_dir}: the path to the folder that saves DR checkpoints;
	\item \texttt{-tokenizer\_name\_or\_path}: the path or name to the Huggingface tokenizer;
	\item \texttt{-q\_max\_len}: the maximum number of query token;
	\item \texttt{-p\_max\_len}: the maximum number of document token;
	\item \texttt{-qrel\_file}: the path to the TREC format qrel file;
	\item \texttt{-write\_run}: whether to write run files to disk;
	\item \texttt{-output\_dir}: the path to the folder where to save run files;
	\item \texttt{-max\_num\_valid}: the total number of checkpoints need to be validated;
	\item \texttt{-logging\_dir}: the path to folder where saves Tensorboard logs;
	\item \texttt{-metrics}: list of evaluation metrics, such as MRR@10, for reporting checkpoints performance;
	\item \texttt{-report\_to}: list of loggers to report, such as tensorboard or wandb;
\end{itemize}

Asyncval also supports arguments that are compatible with Huggingface such as \texttt{-batch\_size} for setting the encoding batch size and \texttt{-fp16} for trading-off precision and speed. A list of more arguments that can be passed to Asyncval, along with their description, can be found in the repository's README file~\footnote{https://github.com/ielab/asyncval}. After launching the command, Asyncval will listen to the folder indicated by \texttt{-ckpts\_dir} and automatically validate a new model checkpoint whenever saved into the folder. The evaluation metric scores of each checkpoint will be reported to the loggers specified with \texttt{-report\_to}. Currently, Asyncval supports the commonly used \texttt{tensorboard} and the cloud-based logger \texttt{wandb}~\footnote{https://wandb.ai/site}. 

To further speed up checkpoint validation, we also provide a feature to sample from the full corpus a subset, thus to avoid encoding all corpus passages for each checkpoint. The subset is sampled based on a given run file of the validation query set, such as the BM25 results, and a TREC standard qrel file which provides gold (relevant) passages for each query in the validation set. To do this, the following command should be ran:
\begin{lstlisting}[language=bash, label={code:avg}]
python -m asyncval.splitter \
		--candidate_dir str \
		--run_file str \
		--qrel_file str \
		--output_dir str \
		--depth int
\end{lstlisting}
where \texttt{-candidate\_file} is the path to the folder where the pre-tokenized corpus JSON files are saved; \texttt{-run\_file} is the path to the run file; \texttt{-qrel\_file} is the path to the TREC qrel file;  \texttt{-output\_dir} is the path to the folder in which to save the JSON file for the subset; and \texttt{-depth} is the number of top passages to keep for each query in the run file. For example, setting \texttt{-depth} to 100 means that only the top 100 passages for each query are kept. This trades-off validation accuracy for speed. 

Although Asyncval requires an extra GPU to perform training and validation in parallel at training time, the tool can still be used when one GPU is available. In this setup, all model checkpoints are saved during training and Asyncval is ran after the training is complete to automatically validate all saved checkpoints. While this does not provide the time savings from the parallelisation of the training and validation processes, it does spare researchers the effort of coding the validation routine. An additional benefit of using Asyncval in a single-GPU setup is that the corpus splitting feature provided by the tool can still be used, thus making the model validation process more efficient without sensible loss of fidelity.

\section{Demonstration Use Case} \label{sec:results}

Next, we demonstrate a typical use case for Asyncval. For this, we use  Tevatron~\cite{Gao2022TevatronAE}~\footnote{https://github.com/texttron/tevatron} to train a DR model and generate checkpoints during training for testing  Asyncval. We follow the MS MARCO passage DR training example code provided by Tevatron~\footnote{https://github.com/texttron/tevatron/tree/main/examples/msmarco-passage-ranking}, which trains a BERT-based DR model for 100,000 updates and saves checkpoints every 10,000 updates (thus resulting in 10 checkpoints in total). The whole training takes around 8 hours on a Tesla V100 GPU: this leaves a time gap between each checkpoint of around 0.8 hours. For Asyncval, we test the full corpus retrieval (8,841,823 unique passages), subsets of corpus constructed using a BM25 run file with cutoff of 100 (610,411 unique passages) and 1,000 (3,824,631 unique passages), and subsets of the corpus constructed using TCT-ColBERTv2~\cite{lin-etal-2021-batch} with cutoff 100 (641,956 unique passages) and 1000 (4,009,400 unique passages)~\footnote{We use the Pyserini~\cite{Lin2021pyserini} implementation for both BM25 and TCT-ColBERTv2.}. We use MRR@10 as the evaluation metric to validate checkpoints.


The results are shown in Figure~\ref{fig:fig2}(left). The MRR@10 score  increases during training: the late DR checkpoints are better than the early ones. The validation score trends obtained with the validation subset sampling strategies to speed up validation are the same as the trend produced when validating on the whole MS MARCO validation set (purple line); however, these strategies appear to always overestimate MRR@10 effectiveness compared to the correct value. 
The MRR@10 scores obtained when using subsets induced by TCT-ColBERTv2 (orange curves) are closer to the scores obtained on the full corpus -- curves almost overlap when using the top-1,000 cutoff.
This suggests that if a validation subset is induced from a strong DR baseline, then the estimation of MRR@10 score has higher fidelity.


Asyncval also logs the time required for each checkpoint validation; thse are reported in Figure~\ref{fig:fig2}(right). 
Validating a checkpoint using the full corpus takes more than 2 hours: in this case, the validation time required for each checkpoint is more than the time required to create an individual checkpoint (0.8 hours). When the sampling strategy is used with cutoff 1,000, validation time becomes $\approx1$ hour which is similar to the time required by the training. The fastest validation time is reported for subsets with a cutoff of 100: this only takes $\approx10$ minutes to validate a checkpoint, and each validation thus terminates before the next checkpoint is generated. We note that although a small cutoff is faster and often enough for selecting the ``right'' checkpoint from the current training run, the high fidelity brought by the larger cutoffs is beneficial if researchers are interested in comparing their checkpoints to results from the DR literature.


\section{Conclusions} \label{sec:conclusion}
The validation of dense retrievers checkpoints is harder than for other deep learning models, because it requires a full corpus re-encoding for each checkpoint created. 
We created Asyncval to simplify and provide solutions to speed up this DR validation process.
With Asyncval, researchers can decouple DR checkpoint validation and training, so that validation can be ran on a separate GPU, and they can use the implemented corpus subset sampling strategies to speed up validation without loss of fidelity.
We have demonstrated the use of Asyncval in the context of training a standard dense retriever on the MS MARCO v1 passage dataset -- the procedure and code for other dense retrievers is similar. We have also already used Asyncval to validate new dense retrievers with a different training routine~\cite{zhuang2022characterbert}, demonstrating the versatility of the tool. 
Asyncval is fully open-sourced; code and examples to reproduce the results reported in this paper can be found at \url{https://github.com/ielab/asyncval}.




\bibliographystyle{ACM-Reference-Format}
\bibliography{sigir2022-asyncval}


\end{document}